\documentclass[lettersize,journal]{IEEEtran}
\usepackage[numbers]{natbib}
\usepackage{amsmath,amsfonts}
\usepackage{algorithmic}
\usepackage{array}
\usepackage[caption=false,font=normalsize,labelfont=sf,textfont=sf]{subfig}
\usepackage{textcomp}
\usepackage{stfloats}
\usepackage{url}
\usepackage{verbatim}
\usepackage{graphicx}

\usepackage{amssymb}

\usepackage{xcolor}
\usepackage[figuresright]{rotating}
\usepackage[utf8]{inputenc}
\usepackage[T1]{fontenc}
\usepackage{mathtools} 

\def\BibTeX{{\rm B\kern-.05em{\sc i\kern-.025em b}\kern-.08em
    T\kern-.1667em\lower.7ex\hbox{E}\kern-.125emX}}
\usepackage{balance}
\begin{document}
\title{Distribution System Power-Flow Solution by Hierarchical Artificial Neural Networks Structure}

\author{\IEEEauthorblockN{Arbel Yaniv}
\IEEEauthorblockA{School of Electrical Engineering,
Tel Aviv University,
Tel Aviv, Israel
Email: arbelyaniv@mail.tau.ac.il\\}
\and
\IEEEauthorblockN{Yuval Beck}
\IEEEauthorblockA{School of Electrical Engineering,
Tel Aviv University,
Tel Aviv, Israel}}


\maketitle

\begin{abstract}
In this paper, a new method for solving the power flow problem in distribution systems which is fast, parallel, as well as modular, straightforward, simplified and generic is proposed. This approach is based on a hierarchical construction of an ANNs tree. The power system is divided into multiple clusters, with a modular architecture. For each cluster an ANN is constructed, were the ANNs of the different clusters are organized in a hierarchical manner in which the data from a lower-level layer is fed into an upper layer in accordance with the electric correlation between the clusters. The solution time is fast as it is based on the neural networks predictions and also enables parallel computing of all clusters in any given layer.
The various clusters have a uniform designed single-hidden-layer ANNs, thus providing a straightforward, simple and generic architectural implementation. 
The suggested methodology is an important milestone for bypassing power flow classical methods and introducing a novel machine learning based approach.
The solution for three-phase unbalance IEEE-123 system as well as EPRI Ckt5 system are presented. The predictions of the ANNs of the hierarchical structures are compared to the solution as calculated by OpenDSS simulation software, with very promising results.
\end{abstract}


\section{Introduction}\label{sec1}
The integration of technologies of monitoring, controlling and supervision into conventional power grids has been accelerated in recent years, thus transforming them into smart grids. Such technologies are essential in order to enable high penetration of renewable energy sources that are required in order to decrease greenhouse gas emissions. The problem of solving the state of a grid is called "Power-Flow", and it consists of a system of $2·(n-1)$ non-linear equations, where $n$ is the number of nodes \cite{saadat2009power}. A typical distribution system can have thousands of nodes and this set of non-linear equations must be solved numerically. For control and  optimization applications, one needs to solve the power flow (PF) problem many times, as the search space of possible configurations is very large. 

As a result, for real-time control and optimization purposes, the solution time of the power flow problem is a critical factor. Classical numerical solution methods such as Newton-Raphson (NR), Gauss-Seidel and their derivatives \cite{zhang1997modified},\cite{hasan1992load},\cite{rajicic1988modification}, or sequential Forward Backward Sweep (FBS) algorithm \cite{wang2021multi}, which is also suitable for three phase unbalanced distribution systems\cite{ulinuha2007unbalance} and a modified un-sequential scheme has been developed. However, all of the above mentioned methods are too slow for control and real time optimization applications.

The other limitation of such numerical methods is their dependency on the parameters' data which is often not fully available. For example, in order to solve the PF set of equations, the admittance matrix must be known, whereas practically in distribution systems it is typically only partially known. 

{Both of the above mentioned limitations can be mitigated using ML approaches: the training of
neural networks on historical data measurements eliminates the need of the parameters' data, and while the training stage can be long, once it is done, ANNs yield very fast predictions in comparison to numerical approaches.}

{Neural networks are used for various purposes in the context of power systems, among them is to solve the power flow problem. A work of ML for power system operation support as in \mbox{
\cite{donnot2017introducing}}\hspace{0pt}
,  
includes a preliminary study of testing deep neural networks for approximating load-flow of Matpower 30 and 118-bus grids via Tensorflow framework.
}
{In \mbox{
\cite{hu2020physics}}\hspace{0pt}
, a physics-guided neural network is presented. Inspired
by unsupervised and supervised auto-encoders, a framework of neural networks that simultaneously
model PF solvers and rebuilds the PF model is suggested. However, the suggested model is restricted as it requires accurate topology information.
}

Although deep neural networks such as the above mentioned ones are very effective for euclidean data, they are not suitable for processing graph-structured data, such as the power flow problem, as it may be irregular in comparison to euclidean data. This limitation motivated the development of 
graph neural networks (GNNs) \cite{yang2020power}. GNNs capture the dependence in graphs via the distributed computing theory synchronous messagepassing system. The system, however, is very inefficient when dealing with large graphs
when messages need to pass through long
paths.

GNNs have several uses in the context of power systems, such as parameter and state estimations \cite{pagnier2021physics}. A learning model that utilizes the structure of the power grid is proposed in \cite{zamzam2020physics}. 
It is also usefull for power flow and optimal power flow \cite{owerko2020optimal}, \cite{donon2019graph}. However, this model is limited only for power grids where all lines have the same physical characteristics.

Another variation of this approach is graph convolutional neural network (GCN) \cite{bolz2019power}. This generic and data-driven approach approximates the load flow calculations, by learning the loading on each line instead of the actual voltages.
{An un-supervised graph neural solver was implemented in \cite{donon2020neural}, which calculate power flow by minimizing the violation of Kirchhoff’s law at each bus.
}

While GNN has shown excellent results for certain applications, considering its limitations \cite{yaniv2023towards}, it is still far from being a frontrunner for PF-based applications.

In real-world applications, complex and large problems can often be divided into {sub }problems for simplification. One such problem is the classification task, which is generally a multi-class problem \cite{tumer1996analysis}-\cite{ivascu2021optimising}. There, every neural
network is assigned with a task of solving independently one of
these sub-problems.

Another approach for a scaled solution of unbalanced distribution system (DS) uses relaxed
sub-problems of low complexity.
Such algorithm is based on
the relaxation of the non-linear set of equations as conic
constraints with directional constraints over multiple iterations of a second order cone programming (SOCP) 
\cite{jha2021network}.

In this paper the PF problem is solved by dividing the distribution system into clusters. The division is done by means of InfoMap algorithm \cite{InfoMap}. These clusters are organized in a hierarchical structure. Each cluster is implemented by a designated ANN in such a way that each layer of ANNs feeds on its results for the active and reactive powers to the upper layer of ANNs as an input.
Once the system is divided by InfoMap algorithm, each cluster is solved by a separate single-hidden-layer neural network with a uniform design. 
As Infomap is a multi-level algorithm, it provides a wide variety of possible partition schemes, out of which it is possible to choose a partition of the original graph representing the distribution system according to the architecture and performance objectives.

The paper shows the theory of the proposed method as well as a full simulation {examples }on the unbalanced IEEE-123 {and EPRI Ckt5 distribution networks}. The results are shown to have {a MAE }of up to {1.2}\%,
and the computational time is substantially reduced by {at least a magnitude of order }in comparison to the solution by the numerical method as simulated with OpenDSS which is an open-source program that solves unbalanced distribution systems by {the }fixed-point iteration {method } \cite{zhang2020improved}.

The original contribution of this paper is: 1. Hierarchical structure of ANNs, which is inherently modular and constructed by simplified sub problems of the complete DS topology
2. This hierarchical structure of neural networks and layer- wise parallel computing yields fast prediction in comparison to classical numerical methods.
The novel sub-problems ML orchestrate approach of a solver to the PF problem is purely data-driven, namely, there is no need to know any of the underlying physical topology of the power system.
In comparison to other approaches of ANNs-based implementations for the solution of the PF problem in DSs which are characterized by complex architectures as a result of the characteristics of such real power systems, after applying the division algorithm, the utilization of simple, generic and unified design single-hidden-layer neural networks is possible via the hierarchical tree of ANNs' construction, in accordance to the clusters of the complete power system and the relations between them. This structure is also inherently modular, which is an important advantage as one of the limitations of existing neural network implementations is their limited compatibility to the dynamic nature of DSs, which can have numerous switching events causing often topology changes in a single day due to scheduled maintenance, faults and high penetration of renewable energy resources.

\section{Classical numerical power flow solution of distribution grids}
The PF problem is a nonlinear set of \begin{math}2(n-1)\end{math} equations, where $n$ is the number of nodes of the power system. 


As these equations are nonlinear, numerical methods are classically used as a solution method.
The PF set of equations for unbalanced distribution systems, which are common in the united states, includes three sets of equations:
\begin{equation}{\label{q1}
	I_i^{abc} = \sum_{j=1}^{n}Y_{i,j}^{abc} V_j^{abc}\quad i=1,\ldots,n 
}\end{equation}
where: 
\begin{equation}{\label{q2}
	I_i^{abc} = \left[\begin{matrix}I_i^a \cr I_i^b \cr I_i^c \end{matrix}\right],  V_j^{abc} = \left[\begin{matrix}V_j^a \cr V_j^b \cr V_j^c \end{matrix}\right],   Y_{i,j}^{abc} = \left[\begin{matrix}
Y_{i,j}^{aa} & Y_{i,j}^{ab} & Y_{i,j}^{ac} \cr Y_{i,j}^{ba} & Y_{i,j}^{bb} & Y_{i,j}^{bc} \cr Y_{i,j}^{ca} & Y_{i,j}^{cb} & Y_{i,j}^{cc} \end{matrix}\right]
}\end{equation}
{where $I_i^{p}$ is the injected current, $V_i^{p}$ is the complex voltage at bus $i$ for phase $p$, and $Y_{i,j}^{pp'}$ is the element of the admittance matrix connecting buses $i$,$j$ for phase $p$,$p$'. Following (\ref{q1}) and (\ref{q2}), the injected current $I_i^{p}$ is as follows:
}\begin{equation}
	{I_i^{p} = \sum_{j=1}^{n}\sum_{q=a,b,c}Y_{i,j}^{pq} V_j^{q}\quad i=1,\ldots,n 
}\end{equation}
{and the three-phase power flow equations for the unbalanced case are:
}\begin{equation}
	{S_i^p = V_p\sum_{j=1}^{n}\sum_{q=a,b,c}(V_j^{p'})^* (Y_{i,j}^{pp'})^*\quad i=1,\ldots,n 
}\end{equation}
{where $S_i^p$ is the injected complex power at bus $i$ for phase $p$. 
Alternatively, it is possible to }use methods based on symmetrical components \cite{naidoo2018adaptive}.
A PF simulation software that have gained a lot of interest and is used for various applications is OpenDSS \cite{montenegro2015speed}, which is based on a fixed-point iterative method \cite{zhang2020improved}. OpenDSS is an open source software for DS simulations, which is also suitable for unbalanced systems. {It is commonly used as a source for comparison, both for verification as well as for computational comparison purposes for the state-of-the-art solutions in this field 
\cite{mahmoud2020machine}}. The numerical solutions from OpenDSS simulation software are used in this paper as ground truth for the training and the predictions' errors evaluation of each of the ANNs in the ANNs' array structure. 

\section{The proposed hierarchical structure}

\subsection{General}
In this paper we solve the PF problem for DSs by means of hierarchical ANN, namely by dividing it into clusters where each cluster consists of a similar number of nodes. The network division into clusters is presented by means of community detection algorithm as is shown ahead. The conventional PF solver of OpenDSS is used for generating the data for training and testing the ANNs. Therefor, the process of attaining all the data which might take time is only for the training stage and is not counted at the testing stage. An alternative way of attaining the data could have been using historic measured data instead of using a PF program. 

\subsection{ANN array structure implementation}

As the architecture design for a neural network of a distribution power system is a very complicated task due to the large number of nodes in real systems and the complex relations between them, we suggest to divide the power system into multiple clusters of the same order of size. As each cluster is considerably smaller than the complete system, a simple fully-connected neural network (FCNN) can be implemented for each cluster with a uniform choice of hyper-parameters as detailed in the next chapter. The ANNs for the different clusters were organized according to the hierarchical division by Infomap algorithm.

Each ANN is trained and tested with different training and testing sets according to its specific nodes and loads. 

\begin{figure}[htb]
\begin{center}
\centering\includegraphics[width=8cm, height=8cm]{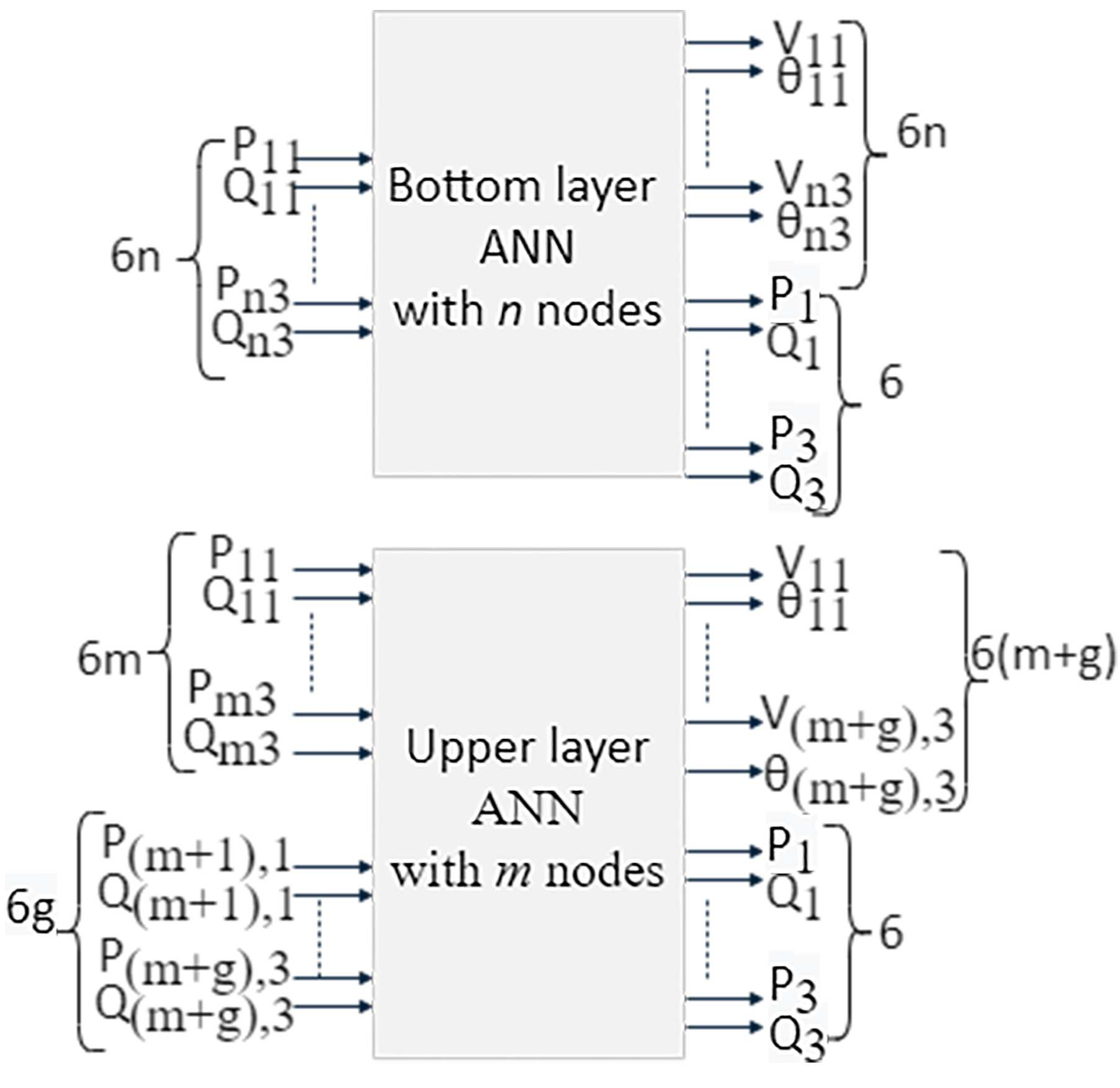}
\caption{{Schematic of the ANNs parameters allocation methodology.}}
\label{colors}	
\end{center}
\end{figure}

Then, each ANN yields the predictions of the voltage amplitudes, phases and correlation preserving parameters. These parameters include the data that needs to be forwarded to the ANNs on the layer above it. This data includes the active and reactive powers at the point of common coupling (PCC) between the layers. The above procedure excludes the top ANN at layer zero  as seen in figure {\ref{colors}} which do not pass on any information due to its location. 

In a hierarchical array of ANNs methodology, there is a combination of the parameters included in an ANN {designed for }a complete power system, and additional parameters which preserve the electrical dependence between the different clusters. The inputs of an ANN for the complete power system are active and reactive powers at each of the loads, as it has only P-Q buses (load buses), and the outputs are the amplitude and phase of the voltage at each of the system's nodes. The additional parameters for a hierarchical ANNs structure are extra inputs or outputs or both according to the location of the cluster in the hierarchy which is dictated according to the partition by the Infomap division algorithm. There are six correlation preserving parameters as inputs/outputs/both (according to the location of the cluster in the hierarchy), namely, {injected active and reactive powers (for each connecting node $i$  there will be $P_{i,1}$,$P_{i,2}$,$P_{i,3}$- the three active powers for each phase and $Q_{i,1}$,$Q_{i,2}$,$Q_{i,3}$- the three reactive powers for each of the phases).}

\begin{figure}[htb]
\begin{center}
\centering\includegraphics[width=7cm, height=8cm]{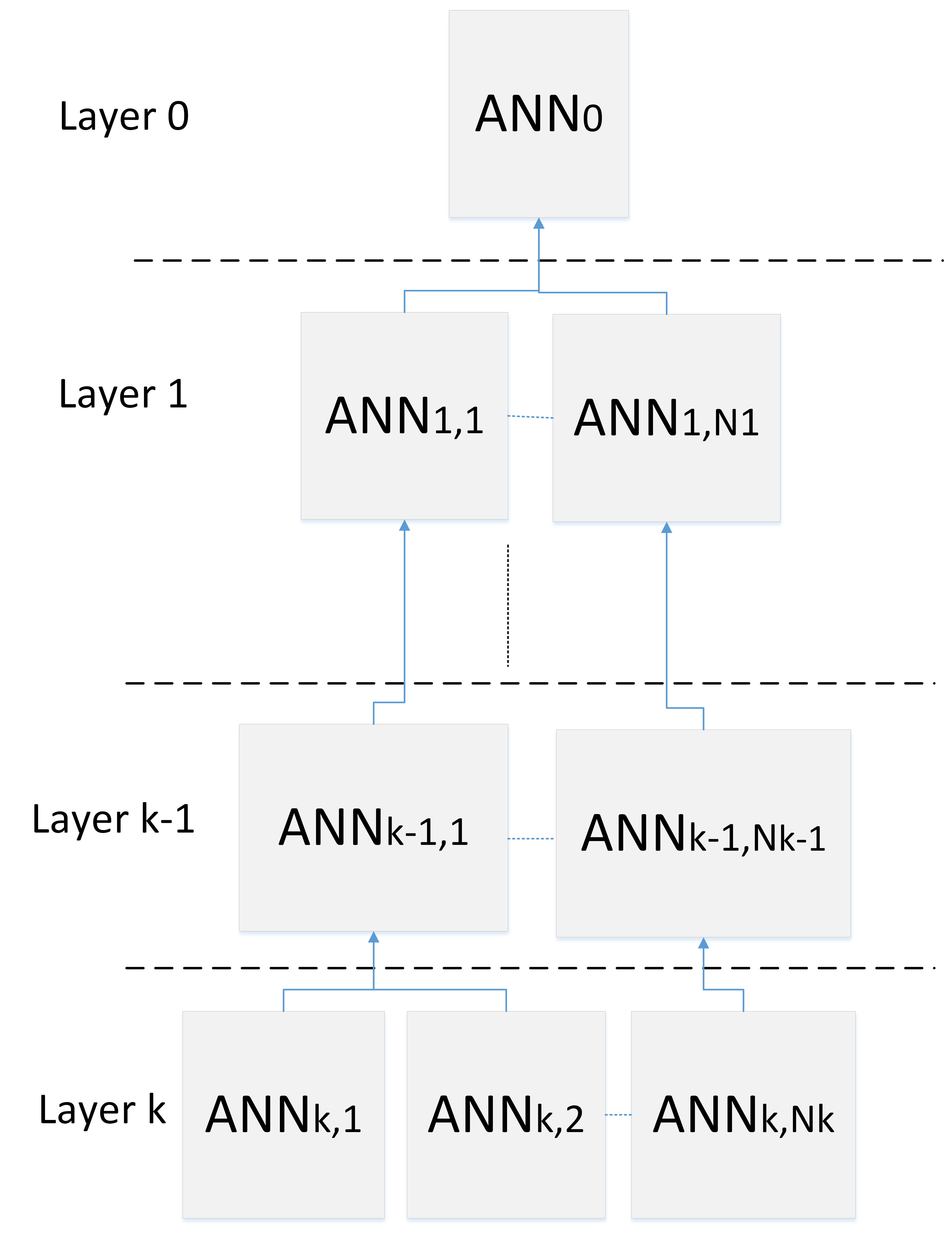}
\caption{{Schematic of an ANNs hierarchical array structure.}}
\label{sch}	
\end{center}
\end{figure}

The criteria for the additional correlation preserving parameters are as follows. Each cluster which has no clusters beneath it in the feeder (leaf), has additional output parameters of the active and reactive power which flows into the head node of that cluster (at each phase). These output parameters are than pass as input parameters to all of the ANNs which belongs to the clusters which are at the next higher layer, thus preserving the correlation of all of the leaf clusters to the cluster above it (its parent). Note that the correlation preserving parameters pass from one ANN to another at the ANN testing stage are all the solutions as predicted by the relevant ANN. In the same manner, clusters at a mid layer, have output parameters of the active and reactive power which flows into the common coupling node (CCN) of that cluster (at each phase), and additional input parameters from the cluster below it (from each of its children). Meaning, in case that a cluster has multiple clusters below it, it will have additional input parameters for each of the CCNs of the clusters below it (child). The only cluster with no additional output parameters is the cluster at the top of the DS (with the head node which is the slack bus of the entire system), as it does not pass any information as there is no clusters above it. This cluster has only additional input parameters of the active and reactive power which flows into the CCN of each of the clusters below it. 
The above mentioned parameters allocation methodology is demonstrated in fig. \ref{colors}: As can be seen, there is a slight difference if the ANN is a bottom layer ANN or an upper one, due to the fact that bottom layer ANN are not fed with data from lower levels. Therefore, for a bottom layer cluster with $n$ nodes, three sets of active and reactive powers will be the input of the cluster, namely, a total of $6n$ inputs. ANNs at upper layers have an inherent feed from a lower layer. For a cluster with $m+g$ nodes, where $m$ is the number of independent nodes and $g$ is the number of nodes that are fed from the lower layer, there will be $6m$ independent inputs and $6g$ inputs that are fed from the lower level. The minimum value of $g$ is one, and for this case there will be only six inputs (three sets of active and reactive power for each phase). The output data will be the voltage amplitudes and angles at all $m+g$ nodes. Both for bottom layer as well as for upper layer clusters (excluding the cluster at the top level which do not feed forward on any information), there are additional six correlation persevering output parameters of the head node of the cluster that will be used to feed the upper-level cluster as explained above.
A schematic of an ANNs hierarchical array structure is shown in Fig. \ref{sch}. It can be seen that the hierarchy is divided into layers ($k+1$ layers in the figure). The head bus is included with the upper ANN at layer 0 which consists of only a single ANN. This ANN will be the last one to be calculated and will be fed by the data of the loads included in that cluster and the data that is fed to this ANN from lower levels. Layer $1$ will have $N_1$ ANNs, layer $2$ will consist of $N_2$ ANNs and layer $k$ will have $N_k$ ANNs accordingly.

\subsection{Community detection algorithm- InfoMap}

As above mentioned, in this paper the algorithm for dividing the DS into clusters is {a }community detection algorithm.

The objective of community detection in power systems is to learn how a network's structure influences the system's behavior by identifying its modular structure with respect to flow of resources. This can be done by exploiting the inference-compression duality.

According to the statistical minimum description length (MDL) principle \cite{grunwald2005advances}, any set of data can be represented by a string of symbols from a finite alphabet, since any regularity in a set can be used to compress it. Hence, this principle can be used to find structures that are significant with respect to how resources flow through networks. 
This also implies that there is a duality between inference to compression of networks. 
This flow can be found according to a communication process in which a sender wants to communicate to a receiver regarding its trajectory. Thus, the trace of the network's flow is represented by a compressed message. 

The InfoMap algorithm is used for the network division and is based on a hierarchical version of the map equation \cite{rosvall2009map}. {The core of Infomap algorithm follows the Louvain method: neighboring nodes are joined into modules, which subsequently are joined into supermodules. The hierarchical rebuilding of the network is repeated until the map equation cannot be reduced further.
Built upon that, Infomap generalizes this search algorithm of the two-level map equation into a multilevel algorithm by a recursive search which operates on a module at any level, where for every split of a module into submodules, the two-level search algorithm is used.
}

Infomap was already used specifically for power grid hierarchical segmentation\cite{marot2017large}, and for guided machine learning (ML) for power grid segmentation for the task of active power management\cite{marot2018guided}.

The multilevel characteristic of Infomap makes it specifically suitable for the hereby proposed approach of array of ANNs, as it is possible to choose the granularity of the partition. In case the first level partition is highly unbalanced, it is possible to continue to the second level partition of the sub-modules, and continue even farther for next level partitions as desired. We hence choose the hierarchical community detection algorithm InfoMap.

\section{ANN unit topology for each cluster}
A common learning architecture is the multi-layer perceptron (MLP) artificial neural network (ANN)
\cite{abid2006accelerating}. It can be represented as a finite directed acyclic graph organized into layers. In this graph, the nodes that do not receive connections from other nodes are referred to as input neurons. Nodes that do not send connections to other nodes are known as output neurons. The remaining nodes that lie between the input and output layers are called hidden neurons.

{Using historical measurements or synthetic databases, such ANNs can be trained in a supervised manner by a training stage where the outputs are given as a part of the database, to approximate the function describing the relation between the inputs and the outputs of the ANN. In the power flow set of equations, the inputs and outputs are the known and un-known electrical parameters respectively. As elaborated in the previous section, in the proposed methodology the power system is divided into clusters and an ANN is assigned to each cluster. Thus, the hyperparameters' selection process for the ANNs assigned to the clusters is elaborated next.
}

\subsection{Model's hyperparmaters}
Uniform design characteristics were implemented for each of the ANNs for each of the clusters.
The ANNs were chosen to be multi-layer perceptron regressors (MLPRs) with a single hidden layer according to the universal approximation theorem \cite{goodfellow2016deep}. According to the theorem, a single hidden layer standard multilayer feed-forward network with a finite number of hidden neurons, is a universal approximator among continuous functions on compact subsets of $Rn$, under mild assumptions on the activation function. As Q(v) and P(v) are continuous functions on compact subsets of $Rn$, the theorem is adequate for the power flow use case under a suitable activation function.
 A long process is involved in the construction of ANNs, which involves both theory and trial and error. Part of the implementation choices were taken according to \cite{yaniv2021state}, where the guidelines for the construction of an ANN architecture for a distribution system are laid out as a preliminary work of the authors. 
 
\begin{figure}[htb]
\begin{center}
\centering\includegraphics[width=8cm,height=3cm]{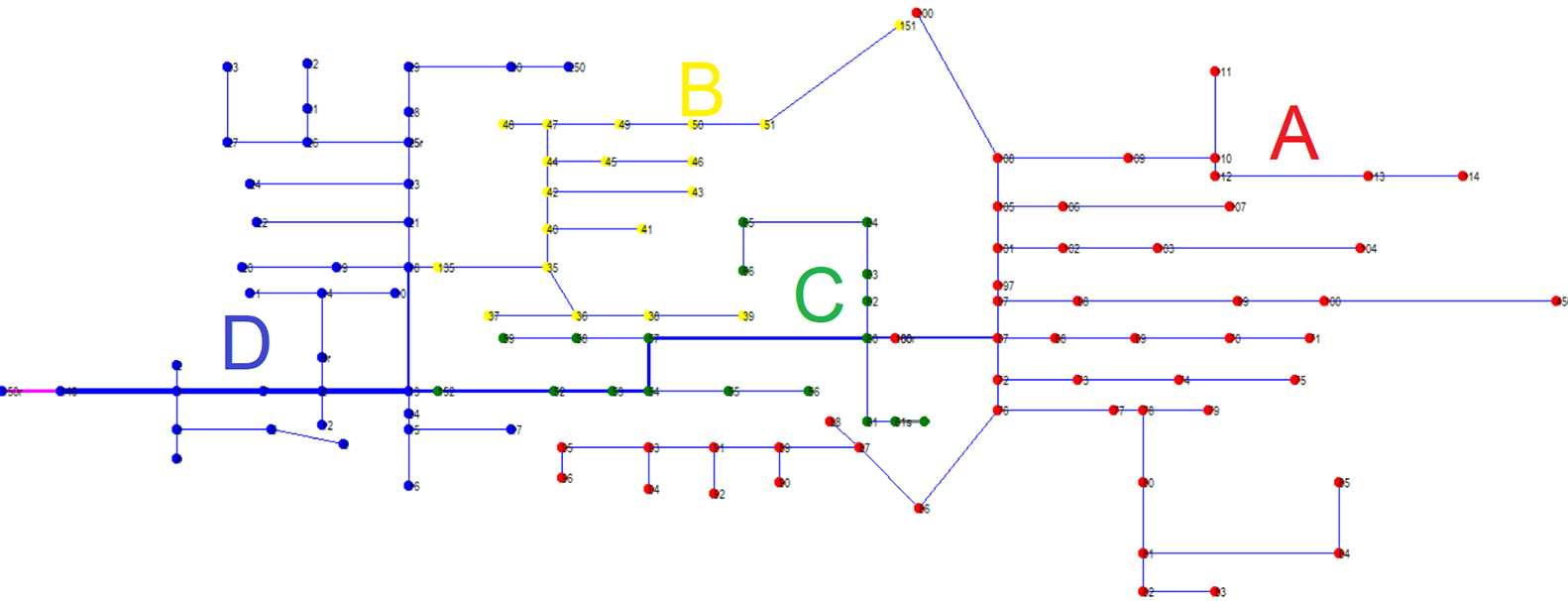}
\caption{{Schematic topology of IEEE-123, divided according to the InfoMap algorithm.}}
\label{ieee123withletters}	
\end{center}
\end{figure}

In order to examine the choices of the various hyperparameters, an automated procedure was done via 
Talos. Talos is a python hyperparameter optimization library for Keras which  allows to configure, perform and evaluate hyperparameter optimization experiments.  
With Talos, numerous experiments were conducted with different combinations of hyper-parameters options. The set of options was constructed by narrowing down relevant values according to the preliminary mentioned processes.
The experiments were conducted on cluster A as depicted in figure \ref{ieee123withletters}, and the parameters' options are detailed in table \ref{talos}. 
More than 15,000 configurations have been tested, to cover a wide range of possibilities.

There was almost a definite division of the configurations performances according to the optimizer, from stochastic gradient descent (SGD) with the highest error,  Root Mean Square Propagation (RMSprop) to adaptive moment estimation (Adam) with the lowest error.
Indeed, Adam was chosen originally as the optimizer for our ANNs. The complete hierarchical architecture was tested with one of the best performing  architectures from the experiment, and achieved similar results to the preliminary chosen architecture. Thus, the simulations provide a quantitative assessment for the hereby chosen design.

\begin{figure}[t]
\begin{center}
\centering\includegraphics[width=\linewidth,height=7cm]{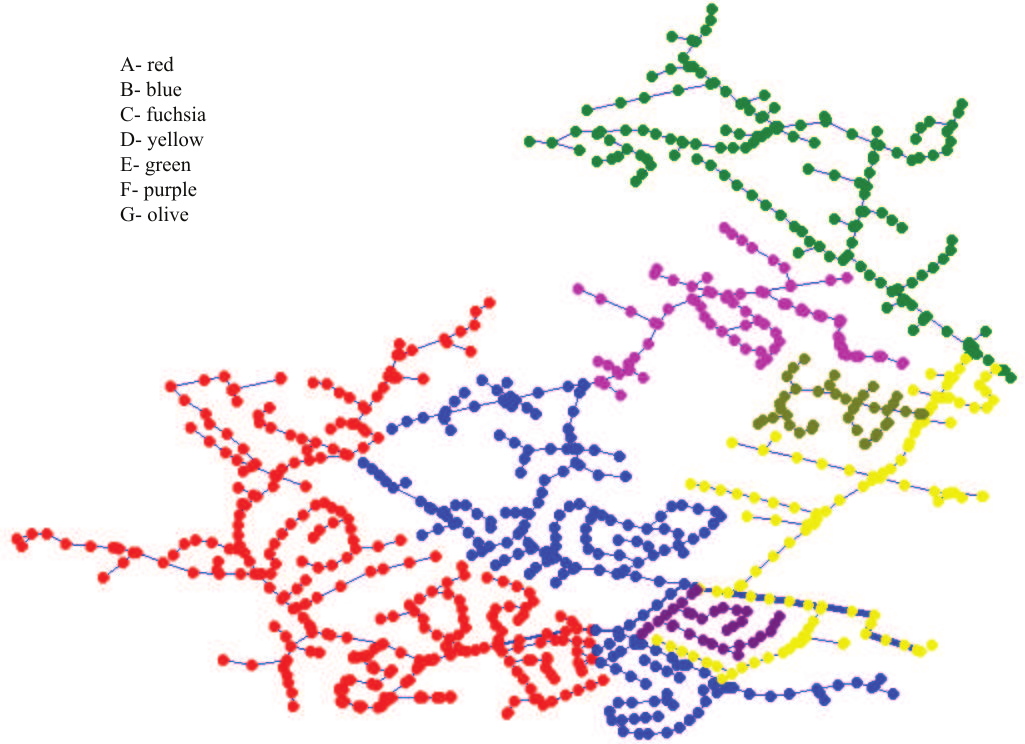}
\caption{{Schematic topology of EPRI Ckt5, divided according to the InfoMap algorithm.}}
\label{ckt5withletters}	
\end{center}
\end{figure}

\begin{figure}[h]
\begin{center}
\centering\includegraphics[width=8cm]{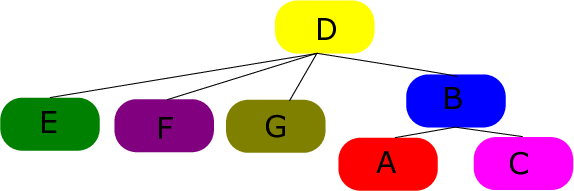}
\caption{{Schematic of the ANN's array structure implementation for EPRI Ckt5.}}
\label{ckt5_scheme}	
\end{center}
\end{figure}
\subsection{Generation of training and testing sets for the ANNs}
Load-shape is a vector that is used to describe many input states of P-Q nodes in a power system. The loadshape is usually normalized and these multipliers are adjoint with the active and reactive power values as specified in the power system's definitions for attaining the values of various types of loads (such as domestic, industrial and commercial). Since for the construction of a database, each input parameter should be assigned with numerous values in a long time series. And, in general, different loads have different values at each point in time, an assignment of different load-shapes to the active and reactive power of the P-Q buses is needed. The multiplication of the load shapes with the active or reactive power definitions yields a vector which represents the behavior of the active or reactive power at that load throughout a hole year. There are different ways to synthetically generate multiple load-shapes out of a single load shape and a few attempts were introduced by adding noise distribution to the load-shape 
\cite{zhou2016residential}- 
\cite{prudenzi2002neuron}. 

The process of generating a substantial amount of load profiles for the various loads is done by using a non-linear companding function such as the well-known \begin{math}\mu\end{math}-law function as commonly used in digital communication \cite{smith1997scientist}.


In this paper, rather than solving a set of nonlinear equations, we approach the problem using supervised ML. The method is based on an ANN, and solves the problem based on a training set composed of the results of many state solutions for the power systems produced by OpenDSS simulation software. It should be mentioned here that the choice of classical algorithm is not important, and can be changed. Since it is used only to generate the database and to verify the results, any suitable algorithm could be used. The time required to generate the data and the convergence time of the OpenDSS simulations are transparent to the ANN, since it is used only at the learning stage.

In order to have supervised learning for the neural networks, a dataset must be created. The dataset includes many inputs and outputs, some of them (80\%) are used for training the ANNs, and some of them (20\%) are used for testing and evaluating the ANNs performance for new unseen inputs. The inputs for each state of the system at a given point in time are the values of active and reactive {power values }at all loads and input correlation preserving parameters. The OpenDSS simulator calculates the solution numerically.{ As the power flow calculations of OpenDSS are based on the fixed point iterative numerical method, it sometimes does not converge to the correct result, as numerical methods are characterized with phenomenons of error accumulation
\cite{liu2018data} and non-convergence \cite{baghaee2017three}. As a result, an outliers removal procedure was used, via a moving median filter with a window size of three samples. Then, }the amplitude and phase of the voltage at each of the nodes and the relevant correlation preserving parameters {are collected and }used as the outputs {for the training and evaluation corresponding to the }different input states.
After the training process of the ANNs is done, the PF solution will be achieved as the predication (output) of the ANNs, instead of a numerical calculation.

The power system is divided by Infomap algorithm  according to the distribution systems connectivity, as it yields from the admittance matrix generated by OpenDSS. A different data-base is generated for each of the clusters according to the algorithm's division, where each database is constructed from the parameters of the buses which belongs to each cluster.

\section{Simulation results}
\subsection{{General}}
In this section, we present the results of the {simulations of two distribution system, } IEEE-123 system {and EPRI Ckt5 system}.
The ground truth data was generated in incorporation with OpenDSS's COM interface \cite{montenegro2015speed}, and was divided into training and testing sets. The error metric used to evaluate the quality of the voltage amplitudes and phases  predictions is the mean absolute error (MAE) {and maximum absolute error (MAXAE)}:
\begin{equation}
MAE=\frac{\sum_{i=1}^{N} |y_i-\hat{y_i}|}{n}
\end{equation}
\begin{equation}
    {MAXAE=\max_{i=1..N} |y_i-\hat{y_i}|
}\end{equation}
where \begin{math}N\end{math} is the number of testing samples, $y_i$ is the  ground truth value according to the OpenDSS simulator results and \begin{math}\hat{y_i}\end{math} is the predicted value according to the ANN.
The error metric for the active and reactive power at the head of the clusters is the relative MAE also known as Mean Absolute Percentage Error (MAPE). Namely, the absolute error is divided by the absolute value of the ground truth value, as the active and reactive power errors are relative to the ground truth value. 
{Correspondingly, the relative MAXAE also known as Maximum Absolute Percentage Error (MAXAPE) is the maximum absolute error divided by the absolute value of the ground truth value.
}For each cluster, the errors were averaged over the {power systems' }buses for 20 consecutive runs with the same training and testing sets.

\subsection{{Simulation results for the IEEE-123 distribution system}}

In this section, we present the results of the OpenDSS simulations as well as the predictions of the ANNs for the IEEE-123 system.

IEEE-123 distribution system \cite{feeders2013ieee} operates at a nominal voltage of 4.16 kV. The topology is shown in Fig. \ref{ieee123withletters}. 

\begin{figure}[h!]
\begin{center}
\centering\includegraphics[width=8cm]{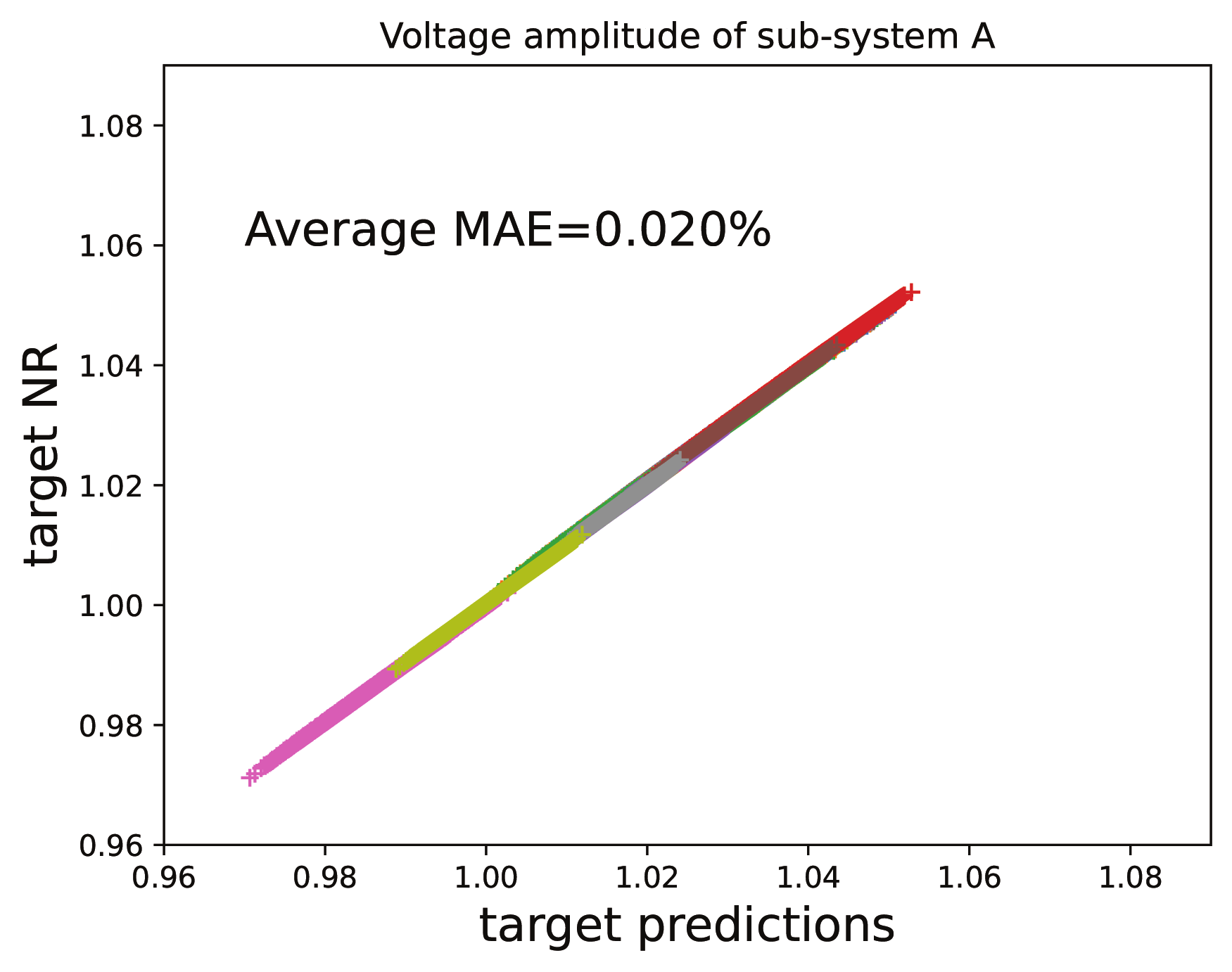}
\caption{{Voltage amplitude of numeric results (NR) calculated via OpenDSS vs. those predicted with the ANN of cluster A of IEEE-123}}
\label{A_4y}
\end{center}
\end{figure}

\begin{figure}[h!]
\begin{center}
\centering\includegraphics[width=8cm]{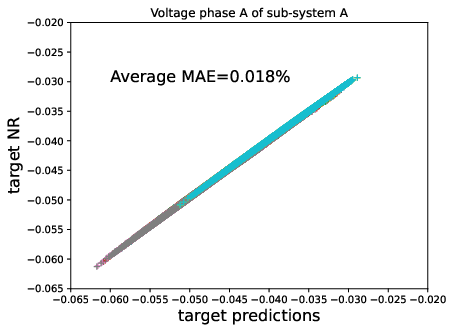}
\caption{{Voltage phase A of numeric results (NR) calculated via OpenDSS vs. those predicted with the ANN of cluster A of IEEE-123}}
\label{A_p1}
\end{center}
\end{figure}

\begin{figure}[h!]
\begin{center}
\centering\includegraphics[width=8cm]{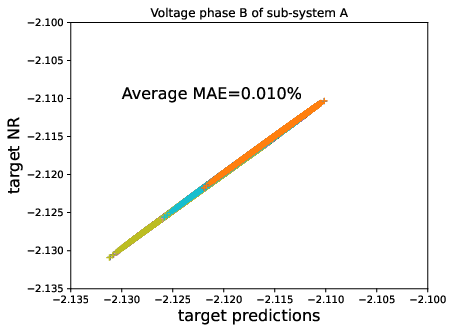}
\caption{{Voltage phase B of numeric results (NR) calculated via OpenDSS vs. those predicted with the ANN of cluster A of IEEE-123}}
\label{A_p2}
\end{center}
\end{figure}

The simulation results for the ANN of cluster $A$ shows a MAE of 0.021\% for the voltage amplitudes, as shown in Fig. {\ref{A_4y}}, where each color represents a different node in the cluster{. Each }point within each color represents a {single }sample from the testing set {at a different point in time.
The results are }0.0189\% for phase $A$ (the 0$^{\circ}$ phase), 0.01\% for phase $B$  (the 120$^{\circ}$ phase shift) and 0.012\% for phase $C$ (the -120$^{\circ}$ phase shift) MAE respectively, as shown in Figs. \ref{A_p1}-\ref{A_p2} respectively.
The relative MAE of the apparent power of the head node of cluster $A$ is 0.407\%.
Similar results were obtained for clusters $B$, $C$ and $D$, and are detailed {in }table \ref{results}.

\begin{table*}[hbt] 
\begin{center}
\caption{ANNs predictions errors of clusters A,B,C and D of IEEE-123 system.} 
\begin{tabular}{ |p{2cm}|p{2cm}|p{2cm}|p{2cm}|p{2cm}|p{2cm}| }
\hline
\cline{2-6} 
Error \% & \textbf{\textit{voltage amplitude MAE}}& \textbf{\textit{voltage phase A MAE}}& \textbf{\textit{voltage phase B MAE}}& \textbf{\textit{voltage phase C MAE}} & \textbf{\textit{S of cluster's PCC MAPE}}\\
\hline
Cluster A & 0.021 &0.0189 &0.01 &0.012 & 0.407\\
\hline
Cluster B & 0.011 &0.01 &0.006 &0.005 & 0.359\\
\hline
Cluster C & 0.014 &0.013 &0.002 &0.009 & 0.381\\
\hline
Cluster D & 0.011 &0.01 &0.005 &0.006 &None\\
\hline
\end{tabular}
\label{results}
\end{center}
\vspace{-0.25cm}
\end{table*}

\begin{figure}[h!]
\begin{center}
\centering\includegraphics[width=8cm]{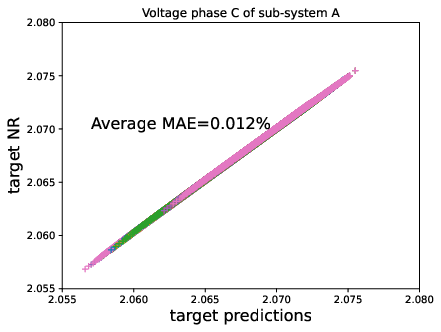}
\caption{{Voltage phase C of numeric results (NR) calculated via OpenDSS vs. those predicted with the ANN of cluster A of IEEE-123}}
\label{A_p3}
\end{center}
\end{figure}

\begin{figure}[h!]
\begin{center}
\centering\includegraphics[width=8cm]{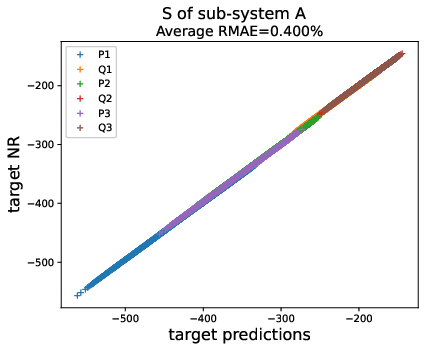}
\caption{{Apparent power of numeric results (NR) calculated via OpenDSS vs. those predicted with the ANN of cluster A of IEEE-123}}
\label{A}	
\end{center}
\end{figure}

\subsection{{Simulation results for EPRI Ckt5 distribution system}}
{In this section, we present the results of the OpenDSS simulations as well as the predictions of the ANNs for EPRI Ckt5 system.
}

{EPRI Ckt5 \cite{ckt5_a} operates at a nominal voltage of 12.47 kV, with a total of 16,310 kVA service transformers. The topology is shown in Fig. \ref{ckt5withletters}. 
}

{The simulation results for the ANN of cluster $A$ shows a MAE of 0.055\% for the voltage amplitudes.
The results are 0.069\% for phase $A$ (the 0$^{\circ}$ phase), 0.088\% for phase $B$  (the 120$^{\circ}$ phase shift) and 0.074\% for phase $C$ (the -120$^{\circ}$ phase shift) MAE respectively.
The relative MAE of the apparent power of the head node of cluster $A$ is 1.065\%.
Similar results were obtained for clusters $B$-$G$, and are detailed at table \ref{results2}.
The results of the MAXAE and MAXAPE are also detailed in table \ref{results2}, where clusters B,D and E consists of nodes connected to all three phases, and clusters C,F and G consists of nodes connected to phase B,B and C correspondingly.  {The results for the voltages of all clusters are consistently small, as well as just over 1\% MAE for the apparent power at the PCC. The MAXAE results are less than 1.508\% for all the clusters' voltages and less than 10\% for the PCC apparent power MAXAPE. These results are common in the field, and are consistent with reported results of other works such as in \cite{donon2020neural},\cite{pham2022neural},\cite{donon2019graph},\cite{donnot2017introducing}. It should be noted that the results in these works were obtained for a distribution systems with around 100 buses while in this paper similar errors were derived for a network of over 3,000 nodes.
}

\begin{table*}[hbt] 
\begin{center}
\caption{{ANNs predictions errors of clusters A-G of EPRI Ckt5 system.}} 
\begin{tabular}{ |p{2cm}|p{2cm}|p{2cm}|p{2cm}|p{2cm}|p{2cm}| }
\hline
\cline{2-6} 
{Error \% }& \textbf{\textit{{voltage amplitude MAE | MAXAE}}}& \textbf{\textit{{voltage phase A MAE | MAXAE}}}& \textbf{\textit{{voltage phase B MAE | MAXAE}}}& \textbf{\textit{{voltage phase C MAE | MAXAE}}} & \textbf{\textit{{S of cluster's PCC MAPE | MAXAPE}}}\\
\hline
{Cluster A }& {0.055 | 0.239 }&{0.069 | 0.265 }&{0.088 | 0.339 }&{0.074 | 0.288 }& {1.06 | 5.91 }\\
\hline
{Cluster B }& {0.046 | 0.234 }&{0.067 | 0.295 }&{0.079 | 0.332 }&{0.072 | 0.316 }& {1.081 | 6.38}\\
\hline
{Cluster C }& {0.075 | 0.311 }& {None }&{0.091 | 0.35 }& {None }& {1.175 | 8.356}\\
\hline
{Cluster D }& {0.246 | 0.36 }&{0.438 | 1.357 }&{0.48 | 1.496 }&{0.486 | 1.508 }&{1.76 | 9.98}\\
\hline
{Cluster E }& {0.056 | 0.239 }& {0.079 | 0.31 }&{0.078 | 0.301 }&{0.082 | 0.318 }&{1.129 | 6.962}\\
\hline
{Cluster F }& {0.054 | 0.368 }& {None }&{0.083 | 0.572 }& {None }& {1.06 | 6.931}\\
\hline
{Cluster G }& {0.06 | 0.249 }& {None }& {None }&{0.084 | 0.355 }&{1.194 | 7.574}\\
\hline
\end{tabular}
\label{results2}
\end{center}
\vspace{-0.75cm}
\end{table*}

\section{Computational considerations}

Among the advantages of the proposed hierarchical ANN tree structure, as a solution approach for the PF problem in distribution systems, is the reduced solution time in comparison to classic methods. 
This improvement is necessary due to the advancements in DSs and the desire to implement control of assets in real time applications (real time here is in the range of minutes in practical cases).

The classical solution based on power flow solver (OpenDSS), as well as the ANNs were tested on a single computer for comparison. The computer is 8th generation i7 1.8GHz 8GB RAM Intel processor. It should be mentioned here that the importance is the comparison of the difference in the execution time and not the actual numbers. This is important since in industrial applications the global time can be significantly reduces by using more advanced and fast computers as well as parallel computing.  

The computational time of an array tree structure (ATS) is:
\begin{equation}\begin{split}\label{time_eq}
	t_{ATS} =
	& max \left ( \sum_{l=1}^{L_{path-i}} max  \left ( t_j \right ) j=1 \ldots N_s \right ) i=1...P
\end{split}\end{equation}
Where $P$ is the number of paths, $L_{path-i}$ is the number of levels in path $i$ and $N_s$ is the number of clusters in each level, $l$, according to the InfoMap community detection algorithm's division.

\subsection{{Computational results for the IEEE-123 system}}
The time it took for each ANN to predict the testing set of IEEE-123 system is detailed in table \ref{times}, averaged over 20 runs. As can be seen, the testing time of the array tree structure takes 0.022 seconds.

The solution time of the testing set for IEEE-123 via OpenDSS's python COM interface takes  1.3 seconds.
Hence, the solution time via the array structure of ANNs is improved by a factor of 50 in comparison to the classical approach via the OpenDSS simulation software.

\begin{table}[h]
    \begin{minipage}{0.5\linewidth}
      \caption{IEEE-123 ANNs' predictions times}
      \centering
        \begin{tabular}{|c|c|}
            \hline
            Cluster name & \textit{Testing time [s]}\\
            \hline
            Cluster A & 0.012 \\
            \hline
            Cluster B & 0.005 \\
            \hline
            Cluster C & 0.004 \\
            \hline
            Cluster D & 0.006 \\
            \hline
        \end{tabular}
        \label{times}
    \end{minipage}
    \begin{minipage}{0.45\linewidth}
      \centering
        \caption{Ckt5 ANNs' predictions times}
        \begin{tabular}{|c|c|}
            \hline
            Cluster name & \textit{Testing time [s]}\\
            \hline
            Cluster A & 7.005 \\
            \hline
            Cluster B & 4.743 \\
            \hline
            Cluster C & 1.841 \\
            \hline
            Cluster D & 2.367 \\
            \hline
            Cluster E & 3.474 \\
            \hline
            Cluster F & 0.388 \\
            \hline
            Cluster G & 0.246 \\
            \hline
        \end{tabular}\label{times2}
    \end{minipage}
\end{table}

\begin{table*}[h]
\begin{center}
\caption{ANN's hyper-parameters options for Talos automated experiments} 
\begin{tabular}{ |p{5cm}|p{2cm}|p{2cm}|p{2cm}|p{2cm}|p{2cm}| }
\hline
\cline{2-6} 
Hyper-parameter & \textbf{\textit{first option}}& \textbf{\textit{second option}}& \textbf{\textit{third option}}& \textbf{\textit{forth option}}&\textbf{\textit{fifth option}}\\
\hline
number of neurons in the hidden layer & eq. 10 from \cite{yaniv2021state}  &eq. 10 from \cite{yaniv2021state} divided by 2 &eq. 10 from \cite{yaniv2021state} divided by 4 &eq. 10 from \cite{yaniv2021state} multiplied by 2 &eq. 10 from \cite{yaniv2021state} multiplied by 4\\
\hline
number of hidden layers & 0&1 &2 &3& \\
\hline
batch size & 32 &64 &128 &256 &\\
\hline
epochs & 50 &100 &150 &200 &\\
\hline
optimizer & SGD &Adam &RMSprop && \\
\hline
activation function of the input layer & linear &relu &tanh&sigmoid &\\
\hline
activation function of the input layer & linear &relu &tanh&sigmoid& \\
\hline
activation function of the input layer & linear &relu &tanh&sigmoid &\\
\hline

\end{tabular}
\label{talos}
\end{center}
\end{table*}

 \subsection{{Computational results for EPRI Ckt5 system}}
The time it took for each ANN to predict the testing set of EPRI Ckt5 system is detailed in table \ref{times2}, averaged over 20 runs. According to equation \ref{time_eq}, the testing time of the array tree structure takes 11.748 seconds.

The solution time of the testing set for EPRI Ckt5 via OpenDSS's python COM interface takes  100.062 seconds.

Hence, the solution time via the array structure of ANNs is improved by a magnitude of order in comparison to the classical approach via the OpenDSS simulation software.

\section{Conclusions and Discussion}
In this paper, a PF methodology by means of a hierarchical array of ANNs was developed. The paper also describes considerations for constructing the appropriate uniform ANN design. The methodology is demonstrated and simulated for the unbalanced IEEE-123 system {as well as EPRI's large-scale Ckt5 system}. 
The discussion of the various considerations and conditions that led to the uniform design of the ANN was also empirically shown to be at least locally optimal through a massive amount of experiments via the hyper-parameter optimization library Talos. 

{The }tree-like graph topological structure of distribution systems is utilized for a hierarchical distributed approach {which }is layed-out, including an appropriate division algorithm for the power system and the detailing of the required parameters for the description of each cluster and for the preservation of the electric information of the related clusters.
The error performance of the results predicted by the trained ANNs tree array are assessed via the comparison to the results obtained from the numerical PF simulation software OpenDSS. The results support the method and theory that is developed in this paper.
The proposed model's performance is shown to be as good as the PF problem's results of much more complicated architectures such as graph neural networks, without the redundant inherent complexity characterizing other deep networks designs \cite{yang2020power}. The computational complexity which is a crucial factor in adopting the suggested ANN approach, is shown to reduce the time to get a result by {at least magnitude of order: a factor of 50 for the IEEE-123 system and a magnitude of order of EPRI Ckt5. }This is important since in real time optimizations, it happens that PF must be performed numerous times for covering a large search space as a result of the large amount of controllable elements in modern smart grids. This improvement enables a result in sufficient time for making a control command.  

The hereby suggested approach offers a basis for a proper operation capabilities of DSs. Massive improvement is required in collecting or generating a quality and generalized database for the continuance of artificial intelligent based applications for smart grids. Smart grid transformation, including high penetration of distributed renewable energy sources towards reducing CO2 emissions, requires a verity of control, monitoring, and supervision technologies. For this purpose, the development of such tools to solve and optimize the system state in real-time are a necessity.

\bibliographystyle{elsarticle-num}
\bibliography{refs}

\end{document}